# Effect of exchange electron-electron interaction on conductivity of InGaAs single and double quantum wells in ballistic regime

**S.V. Gudina[1], Yu.G. Arapov[1], V.N. Neverov[1], A.P. Savelyev[1], S.M. Podgonykh[1,2],**

**N.G. Shelushinina[1], M.V.Yakunin[1,2]**

[1] M.N. Miheev Institute of Metal Physics of Ural Branch of Russian Academy of Sciences, 620137 Yekaterinburg, Russia
[2] Ural Federal University named after B.N. Yeltsin, 620002 Yekaterinburg, Russia

E-mail: svpopova@imp.uran.ru

## Abstract

We report an experimental study of quantum conductivity corrections for two-dimensional electron gas in a GaAs/InGaAs/GaAs single and double quantum wells in a wide temperature range (1.8 – 100) K. We perform a comparison of our experimental data for the longitudinal conductivity at zero magnetic field to the theory of interaction-induced corrections to the transport coefficients. In the temperature range from 10 K up to (45 – 60) K, which covers the ballistic interaction regimes for our samples, a rather good agreement between the theory and our experimental results has been found.

## 1. Introduction

In the early 1980s, a number of experimental works on the temperature dependence of the resistivity $\rho$ (with $d\rho/dT > 0$) in n-type silicon inversion layers for a fairly wide temperature range where $\frac{k_B T \tau}{\hbar} \geq 1$ ($\tau$ is the transport mean free time) has appeared [1-5]. The authors of [1-5] strongly imply that this temperature-dependent scattering is not of phonon origin but is due to electron-electron scattering affected by the disorder (impurities or dislocations).

These studies initiated theoretical works of Stern [6] and of Gold and Dolgopolov [7] in which temperature dependence of mobility for silicon inversion layers has been explained by the temperature dependence of the screening function for elastic scattering. The concept of "temperature-dependent screening", used by Stern [6] and by Gold and Dolgopolov [7], is based on the property of the screening singularity for the Fermi gas (an analogue of the Kohn effect in the spectrum of lattice vibrations [8]), that is on a feature of the screening parameter as a function of wave vector $q$ at $q = 2k_F$.

The temperature smearing of the Fermi step introduces an effective temperature dependence in the screening parameter. The "temperature-dependent screening" is embodied in the temperature dependence of conductivity when the elastic scattering of an electron at the electron density, shielding the impurity, is considered.

The numerical self-consistent calculations of the temperature dependence of mobility carried out by Stern [6] for silicon inversion layers show that the wave-vector and temperature dependence of screening contribute to a temperature-dependent part of the scattering rate, which increases approximately linearly with temperature from 0 to 40 K.

Gold and Dolgopolov [7] presented analytical results for the low-temperature-dependent conductivity $\sigma(T)$ of the two-dimensional electron gas in the presence of charged impurity scattering. Considering the temperature dependence of the screening function for elastic scattering, the anomalous linear temperature dependence of $\sigma(T)$ with characteristic density dependent coefficients has been found.

The results of [6, 7] were reformulated in the work of Zala et al.[9] in the general form

$$\delta\sigma = -\frac{e^2}{\pi\hbar}\left(\frac{k_B T\tau}{\hbar}\right)f(r_s) \qquad (1.1)$$

and it was pointed out that (1.1) corresponds to the ballistic regime, $kT\tau/\hbar \gg 1$ (since the scattering on a single impurity is considered). Here $f(r_s)$ is a positive function of the gas parameter of the system, $r_s$, and thus the equation (1.1) predicts always metallic sign of the interaction correction.

As is known, the presence of a Fermi step leading to the above-described screening features in q-space, causes a modulation of electron density $\delta\rho(r)$, shielding the impurity, in real r-space (Friedel oscillations) [8]. In 2D-case [10] it can be written as

$$\delta\rho(\vec{r}) = -\frac{\nu\lambda}{2\pi r^2}\sin(2k_F r) \qquad (1.2)$$

Here $r$ denotes the distance to the impurity, $\lambda$ being its potential in the Born approximation, $\nu$ is the 2D free electron density of states, $k_F$ is the Fermi momentum. Taking into account the electron-electron interaction (EEI) one finds additional scattering potential due to the Friedel oscillation. This potential can be presented as a sum of the direct (Hartree) and exchange (Fock) terms [9, 11].

Consideration of a scattering by Friedel oscillations in the language of quantum interference effects allowed Zala et al. [9] to obtain the temperature-dependent corrections to conductivity $\delta\sigma^{ee}$. Let us emphasize that the scattering rate becomes temperature dependent just on account of "temperature-dependent screening" as at finite temperatures, due to a smearing of the Fermi step, the Friedel oscillations should be modified as follows [9]:

$$\delta\rho(\vec{r}) = -\frac{\nu\lambda T^2}{2\pi v_F^2 \sinh^2\left(\frac{rT}{v_F}\right)}\sin(2k_F r) \qquad (1.3)$$

For a single scatterer, i.e. in the ballistic limit, $\frac{k_B T\tau}{\hbar} \gg 1$, Zala et al. [9] have found the linear temperature dependence of the quantum correction, $\delta\sigma^{ee} \sim kT/E_F$, in accord with Refs. [6, 7]. However, in contradiction to the result (1.1) of [6,7], the sign of the slope of Zala's dependence is not universal but depends on the strength of EEI (see formulas below).

Method of consideration used in [9] have allowed to upgrade the Eq. (1.1): firstly, to express the coefficient at $k_B T/E_F$ in $\delta\sigma^{ee}(T)$ for the direct Coulomb (Hartree) EEI through the Fermi-liquid constant $F_0^\sigma$ and, secondly, (which is extremely important) to take into account the exchange (Fock) part of EEI. It should be emphasized that these contributions to $\delta\sigma^{ee}$ have opposite signs: the correction has a "metallic" character, $d(\delta\sigma^{ee})/dT < 0$, at predominance of Hartree contribution and "dielectric" behavior, $d(\delta\sigma^{ee})/dT > 0$, at predominance of the exchange part.

Many experimental works, both old [1-5], and the more recent ones (see, for example, [12-15] and an overview of [16] with references therein; see also the extensive bibliography in [17]) are devoted to the observation of the "metallic" behavior of the resistance of 2D- structures in the intermediate and ballistic regime.

On the other hand, such an important aspect of the theory by Zala et al. [9] as the possibility of the dielectric behavior of conductivity in a wide range of rather high temperatures, remained in fact beyond the attention of researchers. As far as we know, 17 years after the work [9] there are experimental works only for few samples that surely demonstrate this behavior (see [17, 18] and [19-21]).

In [17, 18] an experimental study of transport properties in a low mobility, high density two-dimensional electron gas in AlGaAs/GaAs/AlGaAs quantum well with "dielectric" behavior of conductivity, $d\sigma/dT > 0$, in a wide temperature range $(1.5 - 110)$K. A quantitative agreement between a parameter free description of experimental data for the longitudinal conductivity at zero magnetic field and the Hall coefficient with the theory of interaction-induced corrections [9] has been found in both the diffusive and the ballistic regimes.

In [19] an experimental study of interaction quantum correction to the conductivity of n-type $Al_xGa_{1-x}As/GaAs/Al_xGa_{1-x}As$ quantum well at $T=(1.4-10)$ K for a wide range of the parameter $\frac{k_B T\tau}{\hbar} = (0.03 - 0.8)$ is presented. The electron density in the quantum well have been varied by the gate voltage from $n = 1.7\cdot10^{12}$ cm$^{-2}$ to $n = 7\cdot10^{11}$ cm$^{-2}$. Both the $\ln T$ diffusion and linear in $T$ dielectric-type ballistic contributions of the EEI correction have been found.

An essential (up to 30%) increase of carrier mobility with increasing $T$ in the interval of (10-70) K ($\frac{k_B T\tau}{\hbar}$ 0,4–3,8) was observed by us in GaAs/n-In$_x$Ga$_{1-x}$As/GaAs structure with a double quantum well [20, 21], and it was considered to be just due to the quantum correction from the EEI in the ballistic regime.

In this paper we present the results of a study of 2D InGaAs structures with a single and double quantum wells which exhibit a pronounced dielectric type of resistance in a wide range of temperatures (10 ÷ 75) K under the conditions of intermediate and ballistic regimes. Analysis of experimental data on the basis of Zala et al. formulas [9] allowed both qualitatively and to a large extent quantitatively to explain the observed effects just by the dominant contribution of the exchange EEI to the temperature dependence of the electron mobility.

## 2. Theoretical conceptions

It was discovered by Altshuler and Aronov [22] that the Coulomb interaction enhanced by the diffusive motion of electrons gives rise to a quantum correction to conductivity, which in 2D case has the form

$$\delta\sigma^{diff} \sim (e^2/2\pi^2)\ln\left(\frac{k_B T \tau}{\hbar}\right), \qquad \frac{k_B T \tau}{\hbar} \ll 1 \qquad (2.1)$$

The condition $\frac{k_B T \tau}{\hbar} \ll 1$ under which Eq. (2.1) is derived in [22] implies that electrons move diffusively on the time scale $\hbar/k_B T$ and is termed the "diffusive regime".

Zala $et$ $al.$ [9] have obtained the expression for the quantum correction to the conductivity of electrons in two dimensions due to disorder modified EEI at an arbitrary ratio of $k_B T$ and $\hbar/\tau$ in the whole range of temperatures from the diffusive ($\frac{k_B T \tau}{\hbar} \ll 1$) to the ballistic ($\frac{k_B T \tau}{\hbar} \gg 1$) regime. According to [9] a correction to conductivity in the ballistic limit is due to a coherent scattering of electrons by Friedel oscillation of a single scatterer.

Thus in the full expression for the quantum correction, $\delta\sigma^{ee}$, two contributions can be distinguished: diffusion (terms proportional to $\ln\left(\frac{k_B T \tau}{\hbar}\right)$) and ballistic (terms linear in $\frac{k_B T \tau}{\hbar}$):

$$\delta\sigma^{ee} = \delta\sigma^{diff} + \delta\sigma^{ball}, \qquad (2.2)$$

where

$$\delta\sigma^{diff} = \frac{e^2}{\pi\hbar}\left\{1 + 3\left(1 - \frac{\ln(1+F_0^\sigma)}{F_0^\sigma}\right)\right\}\ln\left(\frac{k_B T \tau}{\hbar}\right), \quad (2.3)$$

which coincides with the result of Altshuler-Aronov [22] and Finkelstein equation [23], and

$$\delta\sigma^{ball} = \frac{e^2}{\pi\hbar}\frac{k_B T \tau}{\hbar}\left\{\left[1 - \frac{3}{8}f\left(\frac{k_B T \tau}{\hbar}\right)\right] + \left[1 - \frac{3}{8}t\left(\frac{k_B T \tau}{\hbar}, \tilde{F}_0^\sigma\right)\right]\frac{3F_0^\sigma}{1+F_0^\sigma}\right\}. \qquad (2.4)$$

Here $F_0^\sigma$ and $\tilde{F}_0^\sigma$ are Fermi liquid interaction constants in the triplet channel and the dimensionless functions $f(x)$ and $t(x;\tilde{F}_0^\sigma)$ describe the cross-over between ballistic and diffusive limits. Their full expressions are given in [9], where it was also pointed out that, for numerical reasons, contributions of scaling functions $f$ and $t$ change the result only by few percents and can be neglected for all the practical purposes.

The term in the first square brackets in (2.4) accounts for the exchange (Fock) EEI and the second term in the curly brackets is Hartree EEI contribution. It should be noted [9] that exchange part of EEI correction (2.4) is positive and Hartree contribution is negative (as $\tilde{F}_0^\sigma < 0$ and $|\tilde{F}_0^\sigma| < 1$), thus the total sign of $\delta\sigma^{ball}$ is determined by the difference of these two contributions.

At the present time (see, for example, Renard $et$ $al.$ [17], Minkov $et$ $al.$ [19]), it is assumed that $\delta\sigma^{ball}$ leads to renormalization of the Drude conductivity $\sigma_D^0$ as a result of the temperature dependence of $\tau$ ($\tau \to \tilde{\tau}(T)$), where $\tilde{\tau}(T) = \tau + \delta\tilde{\tau}_{ee}^{ball}$:

$$\tilde{\sigma}(T) = \sigma_D^0 + \delta\sigma_{ee}^{ball} = \frac{e^2}{\pi\hbar}\frac{E_F}{\hbar} + \delta\sigma_{ee}^{ball} \equiv \frac{e^2}{\pi\hbar}\frac{E_F\tilde{\tau}(T)}{\hbar};$$

$$\tilde{\tau}(T) = \tau\left\{1 + \frac{k_B T}{E_F}\left(1 + \frac{3\tilde{F}_0^\sigma}{1+\tilde{F}_0^\sigma}\right)\right\}. \qquad (2.5)$$

In (2.5) we have neglected the corrections from scaling functions $f$ and $t$ (see Eq. (2.4)).

When the magnetic field $B$ is applied, the expressions for $\sigma_{xx}$ and $\sigma_{xy}$, taking into account the contribution of $\delta\sigma_{ee}$, in terms of the mobility $\tilde{\mu}(E) = e\tilde{\tau}(T)/m$ take the following form [17]:

$$\sigma_{xx}(B,T) = \frac{en\tilde{\mu}(T)}{1+\tilde{\mu}^2(T)B^2} + \delta\tilde{\sigma}_{ee}^{diff}(T); \qquad (2.6)$$

$$\sigma_{xy}(B,T) = en\tilde{\mu}(T)\frac{\tilde{\mu}(T)B}{1+\tilde{\mu}^2(T)B^2}. \qquad (2.7)$$

It is most convenient to determine $\tilde{\mu}(T)$ from the $\sigma_{xy}(B,T)$ dependence, since its expression (2.7) does not include $\delta\tilde{\sigma}_{ee}^{diff}$. If $\tilde{\mu}(T)$ is determined from the experiment, we compare $\tilde{\tau}(T) = m\tilde{\mu}(T)/e$ with the expression (2.5) taking into account the interpolation functions $f\left(\frac{k_B T \tau}{\hbar}\right)$ and $t\left(\frac{k_B T \tau}{\hbar}, \tilde{F}_0^\sigma\right)$ at the last stage.

The purpose of comparing the experiment and the theory is to determine the value of the Fermi-liquid constant $\tilde{F}_0^\sigma$. The obtained value $\tilde{F}_0^\sigma$ should also be compared with the theoretical value (see [9]):

$$\tilde{F}_0^\sigma = -\frac{1}{2}\frac{r_S}{r_S + \sqrt{2}} \qquad (2.8)$$

where a gas constant $r_S$ is the ratio of the energy of EEI and the kinetic energy of the electron (for a strongly degenerate electron gas it is the Fermi energy), $r_S = E_{ee}/E_F$. It is convenient to represent the parameter $r_S$ in the following form:

$$r_S = a_{WZ}/a_B^*, \qquad (2.9)$$

where Wigner-Zeitz radius, $a_{WZ} = (\pi n)^{-1/2}$, depends only on the concentration of electrons, and effective Bohr radius, $a_B^* = \kappa\hbar^2/me^2$, is determined by the parameters of a material (dielectric constant $\kappa$ and effective mass $m$).

Let us separate in (2.5) the temperature-dependent part of $\tilde{\tau}$ as

$$\Delta\tilde{\tau}(T)/\tau = A\frac{k_B T}{E_F}, \qquad (2.10)$$

where

$$A = 1 + \frac{3\tilde{F}_0^\sigma}{1+\tilde{F}_0^\sigma} = \frac{1-4|\tilde{F}_0^\sigma|}{1-|\tilde{F}_0^\sigma|}, \qquad (2.11)$$

Then we have the "dielectric" behavior of $\tilde{\sigma}(T)$ ($d\tilde{\sigma}/dT > 0$) for $A > 0$ ($|\tilde{F}_0^\sigma| < 0.25$; $r_S < \sqrt{2}$ ) and the "metallic" one ($d\tilde{\sigma}/dT < 0$) for $A < 0$ ($|\tilde{F}_0^\sigma| > 0.25$; $r_S > \sqrt{2}$ ), where the relation (2.8) between $\tilde{F}_0^\sigma$ and $r_S$ is used.

From the inequalities $r_s \lesssim \sqrt{2}$ we find the value of the concentration $n = N_{cross}$, which determines the hypothetical transition from the "metallic" behavior of $\delta\sigma_{ee}^{ball}$ to the "dielectric" one with an increasing of $n$. The value of $N_{cross}$ depends strongly on the choice of substance (on the parameters $m$ and $\kappa$). Table 1 shows the calculated values of $N_{cross}$ for some substances. We see that these values vary from $(1.1 \div 1.5) \cdot 10^{11}$ cm$^{-2}$ for InGaAs and GaAs up to $(1.5 \div 3.4) \cdot 10^{12}$ cm$^{-2}$ for Si 2D – structures and for n-type silicon inversion layers Si/SiO$_2$.

| Substance | $\kappa$ | $m/m_0$ | $a_B^*$, nm | $N_{cross} \times 10^{-11}$(cm$^{-2}$) |
|---|---|---|---|---|
| GaAs | 12.9 | 0.067 | 10.2 | 1.5 |
| InGaAs (20% InAs) | 13 | 0.058 | 11.85 | 1.13 |
| Si | 11.5 | 0.19 | 3.2 | 15.3* |
| Si/SiO$_2$ | 7.7 | 0.19 | 2.145 | 34* |

Table 1. Material parameters: dielectric constant $\kappa$, effective mass $m$, effective Bohr radius $a_B^*$, and theoretical estimation of concentration $N_{cross}$ for different substances.

Moreover, the estimates in Table 1 for Si and Si/SiO$_2$ are based on the formulas (2.5), (2.10) which does not take into account the presence of two valleys in Si . The increased degeneracy of the system due to the presence of the two valleys may modify numerical coefficient in triplet term [16, 24]. For weak intervalley scattering it depends on the ratio of the valley splitting $\Delta$ to $k_B T$. The Hartree term in Eq. (2.5) becomes equal to

$$\frac{k_B T}{E_F}\left(1 + \frac{15 F_0^\sigma}{1 + F_0^\sigma}\right) ; \quad A = \frac{1 - 16|F_0^\sigma|}{1 - |F_0^\sigma|} \quad \text{for } \Delta \ll k_B T$$

and

$$\frac{k_B T}{E_F}\left(1 + \frac{7 F_0^\sigma}{1 + F_0^\sigma}\right); \quad A = \frac{1 - 8|F_0^\sigma|}{1 - |F_0^\sigma|} \quad \text{for } \Delta > k_B T.$$

Intensive intervalley scattering will result in Eq. (2.5) with $A$ from Eq. (2.11).

As replacing $3 \rightarrow 7 \rightarrow 15$ in the formula (2.5), concentration, $N_{cross}$, of hypothetical transition to the dielectric behavior of $\delta\sigma_{ee}^{ball}$ in silicon structures increases vigorously.

## 3. Experimental results

### 3.1 Samples

The samples were grown by organometallic vapor phase epitaxy on GaAs semiinsulating substrates at Nizhnii Novgorod Physical-Technical Institute of Nizhny Novgorod University by B.N. Zvonkov. The series of structures with single and double quantum wells n-In$_{0.2}$Ga$_{0.8}$As/GaAs was grown with a transition from a double quantum well (DQW) to a single quantum well (SQW), due to a gradual decrease of the barrier width. Preliminary studies of these samples are presented in [25].

The technological parameters of the structures are shown in Table 2. The structures were symmetrically doped in the barriers by Si ($n_D = 10^{18}$ cm$^{-2}$), the width of the spacer was $d_s = 19$ nm. The effective carrier mass was $m = 0.058 m_0$, where $m_0$ is the free electron mass.

The potential profiles of the studied systems, as a function of the growth direction $z$, were obtained from self-consistent solutions of Schrödinger and Poisson equations [25]. In a double tunnel-coupled quantum well the wave functions of the energy levels of each of the two wells are strongly mixed and form symmetric (S) and antisymmetric (AS) states separated by the tunnel gap $\Delta_{SAS}$, which depends on the parameters of the barrier between the wells (see Table 2).

| Sample | $d_s$, nm | $d_w$, nm | $d_b$, nm | $\Delta_{SAS}$, meV |
|---|---|---|---|---|
| DQW-1 | 19 | 5 | 10 | 3.0 |
| DQW-2 | 19 | 5 | 7 | 7.4 |
| DQW-3 | 19 | 5 | 3.5 | 23.1 |
| SQW | 19 | 10 | 0 | — |

Table 2. Technological characteristics of the samples: $d_s$ is the width of the spacer, $d_w$ is the width of the well, $d_b$ is the barrier width, $\Delta_{SAS}$ is the width of the tunnel gap.

It follows from the calculations that two subbands, both of S and AS states, are filled in samples DQW -1 and DQW -2, and only S subband is filled in sample DQW-3 due to a large tunnel gap, $\Delta_{SAS} > E_F$. There is also only one filled subband (S) in the SQW sample.

We have measured the longitudinal $\rho_{xx}(B, T)$ and Hall $\rho_{xy}(B, T)$ components of the resistivity tensor using dc techniques with currents of $\approx 1$ μA in magnetic fields $B = (0 \div 12)$ T, and at temperatures $T = (1.8 \div 100)$K. Experiments were carried out at the Collaborative Access Center "Testing Center of Nanotechnology and Advanced Materials" of the M.N.Miheev Institute of Metal Physics of the Ural Branch of the Russian Academy of Sciences using Oxford Instruments and Quantum Design setups.

| Sample | $n_t$, $10^{15}$ m$^2$ | μ, m$^2$/V·s | $E_F$, meV | $T_{min}$, K | $\sigma(T_{min}) \cdot 10^{-3}$, 1/Ohm | $k_F l$ |
|---|---|---|---|---|---|---|
| DQW-1 | 2.27 | 1.13 | 9.4 | 75 | 0.66 | 17 |
| DQW-2 | 2.05 | 1.66 | 8.4 | 65 | 0.75 | 19 |
| DQW-3 | 2.35 | 2.56 | 9.6 | 45 | 1.12 | 29 |
| SQW | 2.10 | 1.21 | 8.6 | 70 | 0.49 | 12 |

Table 3. Parameters of the samples: $n_t$ is the total charge carrier concentration, μ is the carrier mobility, $E_F$ is the Fermi energy, $T_{min}$ is the transition temperature from "dielectric" to "metallic" type of conduction.

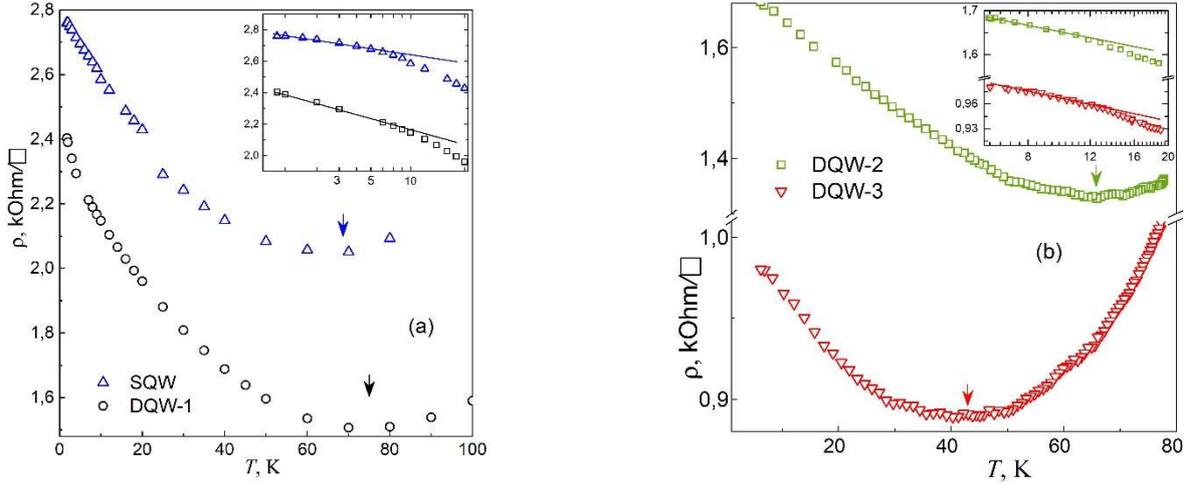

Figure 1. Temperature dependences of the resistivity at $B = 0$ for the samples SQW and DQW-1 (a) and DQW-2 and DQW-3 (b). The arrows show the values of $T_{min}$. Insets: the resistivities $\rho$ for the same samples as functions of $\ln T$ at $T < 20$K. The straight lines are eyeguide lines.

The obtained parameters of the samples are shown in Table 3. The values for $n_t$ and $\mu$ are given at the lowest temperatures of the experiment. The uncertainty in determining the mobility and concentration of the charge carriers does not exceed 3%.

### 3.1 Resistivity at B=0

The temperature dependences of zero-field resistivity $\rho$ for our samples are shown on Fig. 1. A change in the dependence of $\rho(T)$, from "dielectric" behavior ($d\rho/dT<0$) to "metallic"one ($d\rho/dT>0$), was detected for all samples at rather different temperatures $T_{min} = (45 \div 75)$K. On Fig. 1 $T_{min}$ is indicated by arrow, the value of it as well as the conductivity value at $T=T_{min}$, $\sigma(T_{min})$, for each sample are listed in Table 2. A disorder parameter $k_F l$ $\left( \equiv \frac{E_F \tau}{\hbar} \right)$, defined by the relation $\sigma(T_{min}) = \frac{e^2}{h}(k_F l)$, is also presented ($k_F$ being the Fermi momentum and $l$ is the mean free path). Rather high values of this parameter, $k_F l \gg 1$, indicate the good quality of the samples studied.

The low-temperature data for the $\rho(T)$ dependences in a logarithmic scale are presented on the insets of Figs 1a, 1b. It is seen that at $T \lesssim 10$K the dependences of $\rho(T)$ are well described by the logarithmic law that is naturally associated with the contribution of quantum corrections to the conductivity from the weak localization and EEI effects in the diffusion regime.

Unusual is that a decrease in resistance with increasing temperature (the "dielectric" behavior of $\rho(T)$) continues at $T>10$K in a rather wide temperature range up $T = (45 \div 75)$K. The analysis have shown [25], that the behavior of $\rho(T)$ at $T>10$K is caused by the temperature dependence of the carrier mobility, $\mu(T)$.

The "dielectric" range of $\mu(T)$ dependence we presumably have related to the EEI quantum corrections to the conductivity in intermediate and ballistic regimes. In this work we have focused on a quantitative analysis of $\rho(T)$ dependences based on the theoretical concepts of work [9] in order to test this assumption.

The "metallic" decrease of mobility at $T > T_{min}$ is obviously associated with the scattering of carriers on acoustic and optical phonons [25].

### 3.1 A separation of two-carrier contributions

Primarily, we have determined the parameters of charge carries in the samples (concentrations $n(T)$ and mobilities $\mu(T)$) using the magnetic-field dependencies of the longitudinal $\rho_{xx}(B, T)$ and Hall $\rho_{xy}(B, T)$ resistivities measured at various fixed temperatures, which have been recalculated into dependencies of $\sigma_{xx}(B, T)$ and $\sigma_{xy}(B, T)$. The $\sigma_{xx}(B, T)$ and $\sigma_{xy}(B, T)$ for DQW-1 are presented on Fig. 2. Dependencies of $\sigma_{xx}(B, T)$ and $\sigma_{xy}(B, T)$ for other samples have got the same peculiarities of the magnetic-field and temperature behaviour and differ only in numerical characteristics which are systemized in Table 3.

It is convenient to determine concentration and mobility for a sample with one type of carriers from the magnetic-field dependence of the Hall conductivity, $\sigma_{xy}(B)$. It is known that $\sigma_{xy}(B)$ has a maximum at $\mu B=1$, at which its value is equal to $\sigma_D/2$, $\sigma_D = en\mu$ being the Drude conductivity. These relations were used to find both $n(T)$ and $\mu(T)$ for SQW and DQW-3 (table 4a). In the case of two types of carriers (DQW-1 and DQW-2), the mobility determined in such a way gives an effective value of carrier mobility ($\mu_{eff}$) close to the faster carrier one (Fig. 3).

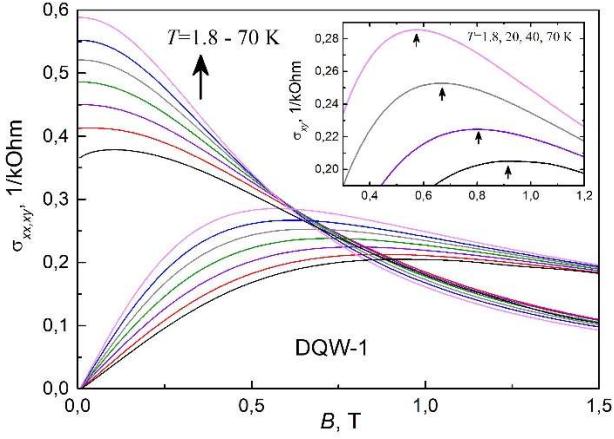

Figure 2. Dependences of the $\sigma_{xx}(B, T)$ and $\sigma_{xy}(B, T)$ measured at various fixed temperatures $T$=(1.8-70) K for DQW-1. Inset shows the parts of $\sigma_{xy}(B, T)$ dependences near the $\mu B$=1 point. Arrows indicate the $\sigma_{xy}(B, T)$ maximum positions.

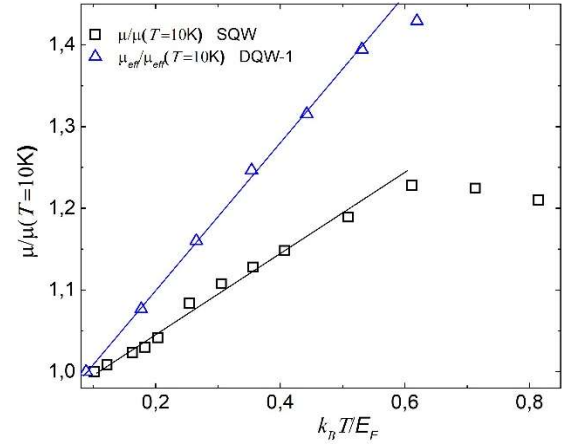

Figure 3. Dependences of the $\mu/\mu(T = 10$ K) on $k_B T/E_F$ for the samples SQW and DQW-1. The straight lines are eyeguide lines.

It is shown in the inset of Fig. 2 that the maximum of $\sigma_{xy}(B, T)$ shifts to the *lower* fields with increasing temperature. This means that the mobility *increases* as the temperature increases and the growth of $\mu(T)$ is essential: $\Delta\mu/\mu(10K) \approx 40\%$ for DQW-1 and $\Delta\mu/\mu(10K) \approx (20-25)\%$ for SQW (Fig. 3) and for the other two samples.

Moreover, a perceptible drop in resistivity with increasing temperature, experimentally observed in our samples in temperature intervals from 10 K to 45-75 K (see Fig. 1), is just due to the mobility increase. The total concentration $n(T)$ remains practically constant for these temperatures as it is demonstrated in Table 4a and on Figs 4a and 5a.

Fig. 3 shows the behavior of $\mu(T)$, found from the position of the $\sigma_{xy}(B)$ maximum, for the selected samples DQW-1 and SQW, but such dependencies are inherent in all the samples. It is seen that these dependencies may with good accuracy be described by the linear law $\Delta\mu(T) \sim k_B T/E_F$ in the range of $0.1 < k_B T/E_F < 0.6$.

We associate the observed unusual behavior of $\mu(T)$ (and $\rho(T)$) in the investigated two-dimensional system with the contribution of quantum conductivity correction due to EEI in the ballistic regime [9] (see Section 2). Within the framework of the Zala *et al.* model [9], natural explanations are found for both the linear dependence of $\Delta\mu$ on $k_B T/E_F$ and the anomalous sign of the derivative $d\mu/dT (> 0)$ as a consequence of the predominant role of the exchange term in EEI correction (see more in Section 4).

Let's emphasize that for systems with two filled subbands, DQW-1 and DQW-2, two types of carriers, the electrons both of S and of AS subbands, participate in the conduction process. An analysis of Hall coefficient $R_H$ and of positive magnetoresistance $\rho_{xx}(B)$ as functions of magnetic fields on the basis of the well-known formulas for two types of carriers

Table 4 (a)

| Sample | $T$, K | $n$, ×10$^{15}$ m$^{-2}$ | $\mu$, m$^2$/V·s | $\tau$ x10$^{-13}$, s | $k_B T\tau/\hbar$ |
|---|---|---|---|---|---|
| SQW | 1.8 | 2.06 | 1.21 | 3.99 | 0.10 |
| | 2.5 | 2.07 | 1.21 | 3.98 | 0.13 |
| | 4.2 | 2.07 | 1.20 | 3.95 | 0.21 |
| | 10 | 2.07 | 1.21 | 3.97 | 0.52 |
| | 20 | 2.04 | 1.25 | 4.14 | 1.1 |
| | 30 | 2.06 | 1.33 | 4.40 | 1.7 |
| | 50 | 2.05 | 1.43 | 4.72 | 3.1 |
| | 70 | 2.01 | 1.47 | 4.86 | 4.5 |
| DQW-3 | 2.6 | 2.35 | 2.61 | 8.61 | 0.29 |
| | 4.2 | 2.48 | 2.60 | 8.59 | 0.47 |
| | 10 | 2.36 | 2.78 | 9.16 | 1.2 |
| | 20 | 2.36 | 2.89 | 9.53 | 2.5 |
| | 30 | 2.29 | 3.01 | 9.94 | 3.9 |
| | 50 | 2.30 | 2.98 | 9.82 | 6.5 |

Table 4 (b)

| Sample | $T$, K | $\mu_1$, m$^2$/V·s | $\mu_2$, m$^2$/V·s | $\tau_1$ x10$^{-13}$, s | $\tau_2$ x10$^{-13}$, s | $\dfrac{k_B T\tau_1}{\hbar}$ | $\dfrac{k_B T\tau_2}{\hbar}$ |
|---|---|---|---|---|---|---|---|
| | | $S$ | $AS$ | $S$ | $AS$ | $S$ | $AS$ |
| DQW-1 | 10 | 1.20 | 1.00 | 3.96 | 3.32 | 0.52 | 0.43 |
| | 20 | 1.27 | 0.91 | 4.19 | 3.02 | *1.1* | 0.79 |
| | 30 | 1.41 | 0.88 | 4.66 | 2.88 | 1.8 | *1.1* |
| | 50 | 1.65 | 0.68 | 5.43 | 2.24 | 3.6 | 1.5 |
| | 70 | 1.77 | 0.67 | 5.83 | 2.21 | 5.4 | 2.0 |
| | | $S$ | $AS$ | $S$ | $AS$ | $S$ | $AS$ |
| DQW-2 | 20 | 1.85 | 1.15 | 6.09 | 3.79 | *0.80* | 0.50 |
| | 30 | 2.02 | 1.05 | 6.65 | 3.46 | 1.7 | *0.91* |
| | 40 | 2.13 | 0.88 | 7.01 | 2.89 | 2.8 | 1.1 |
| | 50 | 2.27 | 0.82 | 7.50 | 2.71 | 3.9 | 1.8 |

Table 4 (a) Carrier concentrations, $n$, mobilities, $\mu$, relaxation times, $\tau$, and parameter $\frac{k_B T\tau}{\hbar}$ at different temperatures for samples SQW and DQW-3. (b) Carrier mobilities, $\mu_i$, relaxation times, $\tau_i$, and parameter $\frac{k_B T\tau_i}{\hbar}$ at different $T$ for S ($i$ =1) and AS ($i$ =2) subbands of samples DQW-1 and DQW-2.

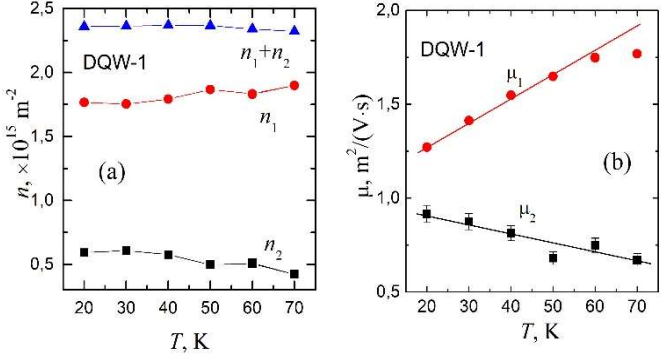

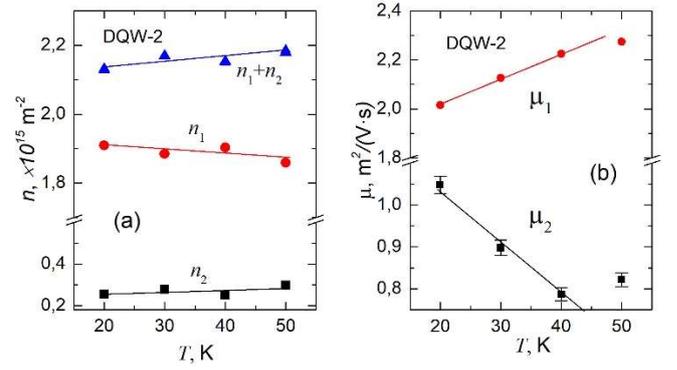

Figure 4. Temperature dependences of the electron concentration (a) and mobility (b) for sample DQW-1: total ($n_t = n_1 + n_2$); in the subbands of symmetric ($n_1$ and $\mu_1$) and antisymmetric ($n_2$ and $\mu_2$) states. The straight lines are eyeguide lines.

Figure 5. Temperature dependences of the electron concentration (a) and mobility (b) for sample DQW-2: total ($n_t = n_1 + n_2$); in the subbands of symmetric ($n_1$ and $\mu_1$) and antisymmetric ($n_2$ and $\mu_2$) states. The straight lines are eyeguide lines.

allowed us to determine the parameters of the electrons in the S and AS states ($n_1$, $n_2$ and $\mu_1$, $\mu_2$, respectively) at different $T$.

The obtained temperature dependences of concentrations and mobilities in the subbands of S and AS states are presented in Fig. 4a,b for DQW-1 and in Fig. 5a,b for DQW-2. It is seen from Figs 4a, 5a the concentrations for each type of carriers, $n_1$, $n_2$, as well as the total electron concentration, $n_t = n_1 + n_2$, are practically independent on temperature.

As for mobilities, the temperature dependencies both of $\mu_1(T)$ and of $\mu_2(T)$ may be rather well approximated by a linear law at temperature intervals (20-60)K for DQW-1 and (20-40)K for DQW-2 (see Fig. 4b, 5b). The most interesting is that the carriers of S - subbands demonstrate the "dielectric" behavior ($d\mu_1/dT > 0$) but the "metallic" behavior ($d\mu_2/dT < 0$) takes place for the carriers of AS - subbands.

The detailed data for mobilities, $\mu$, and relaxation times, $\tau$, as functions of temperature, as well as estimates of the parameter $\frac{k_B T \tau}{\hbar}$ are presented both for samples SQW, DQW-3 with one type of carriers in Table 4(a) and for samples DQW-1, DQW-2 with two types of carriers in Table 4(b). It is seen that a crossover from the diffusion regime to the ballistic one i.e the appearance of the temperature dependence of $\mu(T)$, for investigated samples occurs at $T \approx (10-20)$K for S subbands and at $T \approx (20-30)$K for AS ones where the parameter $k_B T \tau / \hbar \approx 1$ (the values in bold italics in Table 4).

The concentration values found by separation of variables for S and AS subbands in each sample are presented in Table 5.

## 4. Discussion of experimental results

In this Section, we compare the experimental data presented in Section 3 with the theoretical concepts set out in section 2. In the first step, we will make "theoretical" estimates of the parameters $r_s$, $\tilde{F}_0^\sigma$ from Eqs (2.8), (2.9) and of the coefficient $A$ from Eq. (2.11). It will be done for concentrations $n$ found by separation of variables for S and AS subbands in each sample and taking into account the value of $a_B^*$ for the investigated substance InGaAs (see Table 1). The values found for these quantities are presented in Table 5. For comparison the data for the GaAs sample investigated in [17,18] are also given.

As can be seen from Table 5 there is a clear correlation of the sign of $A_{theor}$ with the experimentally observed sign of the derivative, $d\mu/dT$, in the actual temperature interval (see Figs 3, 4b, 5b). This is in accordance with the estimation of $N_{cross}$ presented in Table 1 for InGaAs: $d\mu/dT > 0$ for S- subbands with $n_1 > 1 \cdot 10^{11}$ cm⁻², but $d\mu/dT < 0$ for AS- subbands with $n_2 < 1 \cdot 10^{11}$ cm⁻².

The $d\mu/dT$ sign correlation with the value of $n \lesssim 1 \cdot 10^{11}$ cm⁻² (and with the sign of $A_{theor}$) certainly speaks in favor of the EEI nature for the temperature dependence of $\delta\sigma$ at $T < T_{min}$.

| Sample \ Subband | $n \times 10^{-15}$ m⁻² | $r_s$ | $\|\tilde{F}_0^\sigma\|$ | $A_{theor}$ | $A_{exp}$ | $\alpha$ |
|---|---|---|---|---|---|---|
| GaAs [17,18] | S | 25.4 | 0.35 | 0.1 | 0.62 | 0.82 | 1.3 |
| SQW | S | 2.05 | 1.04 | 0.21 | 0.20 | 0.46±0.02 | 2.3 |
| DQW-3 | S | 2.35 | 0.97 | 0.20 | 0.24 | 0.40±0.01 | 1.7 |
| DQW-2 | S | 1.90 | 1.08 | 0.22 | 0.17 | 0.43±0.03 | 2.5 |
| | AS | 0.25 | 2.98 | 0.34 | -0.54 | -0.16±0.05 | 0.3 |
| DQW-1 | S | 1.75 | 1.12 | 0.22 | 0.15 | 0.57±0.06 | 3.8 |
| | AS | 0.50 | 2.10 | 0.30 | -0.29 | -0.13±0.03 | 0.5 |

Table 5. The parameters $r_s$, $\|\tilde{F}_0^\sigma\|$, $A_{theor}$, $A_{exp}$ and $\alpha = A_{exp}/A_{theor}$, calculated from the values of the electron concentration in the S and AS subbands for the InGaAs samples. Also the data from Ref. 17 and 18 for S subband of the GaAs sample are presented.

For the InGaAs samples studied the parameters of the substance ($m^*$ and $\kappa$) are such that the concentration $N_{cross} = 1.1 \cdot 10^{11}$ cm⁻² falls into a quite accessible range of $n$ value, which allows us to observe experimentally the "dielectric" behavior of the resistivity.

At the same time, in Si/SiO$_2$ systems, mainly because of the large value of $m$ (see Table 1), the oversized concentrations, $n > 3.4 \cdot 10^{12}$ cm$^{-2}$ (or even much higher $n$, taking into account the two valleys), are needed to observe this effect. Apparently, that is why the "metallic" conductivity due to EEI effects in Si/SiO$_2$ is confidently observed up to $n = 1.3 \cdot 10^{12}$ cm$^{-2}$ in [1]; up to $7 \cdot 10^{12}$ cm$^{-2}$ in [2]; up to $3 \cdot 10^{12}$ cm$^{-2}$ in [3]; up to $4.85 \cdot 10^{12}$ cm$^{-2}$ in [4] and up to $7 \cdot 10^{11}$ cm$^{-2}$ in [5].

Only in some cases it was possible to achieve a sign change of d$\delta\sigma_{ee}$/d$T$ with an increase in the electron concentration. Hartstein *et al.* [2] have found a 5% increase of $1/\tau_{ee}$ for $T = (4.2 - 32)$ K when measured the mobility in (001) Si inversion layers with $n = 3 \cdot 10^{12}$ cm$^{-2}$. The measured temperature dependence of $1/\tau$ in [2] was weak at $7 \cdot 10^{12}$ cm$^{-2}$, and $1/\tau_{ee}$ decreased with increasing temperature only at $1.2 \cdot 10^{13}$ cm$^{-2}$.

As for the highest-quality GaAs structures, the estimates lead to a value of $N_{cross}$ (= $1.5 \cdot 10^{11}$ cm$^{-2}$) in a quite accessible range of concentrations (see Table 1). But, paradoxically, it is just the high quality of the samples, which is associated with large spacers and, as a consequence, with the large-scale character of the scattering potential, prevents the observation of the interference EEI contribution in a ballistic regime, $\delta\sigma_{ee}^{ball}$, at $B = 0$ [26, 27]. Due to a small-angle character of the scattering events for a large-scale impurity potential the interaction correction in the ballistic regime at $B = 0$ is suppressed exponentially for the case of smooth disorder as $\delta\sigma_{xx} \sim \exp[-\text{const}((k_B T \tau/\hbar)^{1/2})]$.

In [19] the dielectric behavior of the ballistic EEI contributions to the conductivity of n-type Al$_x$Ga$_{1-x}$As/GaAs/Al$_x$Ga$_{1-x}$As quantum well was observed at $T < 10$ K ($k_B T \tau / \hbar <0.8$) due to not very large electron mobility ($k_B T \tau / \hbar <0.8$) wherein the values of the electron concentration in the sample $n = (0.7-1.7) \cdot 10^{12}$ cm$^{-2}$ well meet the criterion $n > N_{cross}$ for GaAs (see Table 1).

In [17, 18] the "dielectric" behavior of conductivity for GaAs quantum well in a temperature range (20–110) K, which the authors associated with the quantum correction $\delta\sigma_{ee}^{ball}$, could be observed due to the choice of the low mobility, high density system and (one need to add) due to the fact that the electron concentration $n = 2.54 \cdot 10^{12}$ cm$^{-2}$ >> $N_{cross}$ for GaAs.

A systematic shift of the experimental points with respect to the theoretical curve, calculated by Eq. (2.4) for the parameters of their system, was noticed in [17, 18].

The authors explained that by the fact that Zala *et al.* theory [9] describes only the temperature dependence of the conductivity but not the total magnitude of it, disregarding a large $T$-independent interaction-induced contributions. That $T$-independent contribution may lead to renormalization both of the Drude conductivity (background mobility value, μ) and/or of $E_F$ introduced in the theory [9] as the ultraviolet cutoff (see details in [17]).

In the second step, we estimate the experimental value of $A$ ($A_{exp}$) as a coefficient of proportionality between $\Delta\mu(T)/\mu = \Delta\tilde{\tau}(T)/\tau$ and $k_B T/E_F$ (see (2.10)) for each sample. We proceed from the linear interpolation of the experimental data for μ($T$) at (10-20) K< $T < T_{min}$, shown in Fig. 3, 4b, 5b:

$$A_{exp} = \frac{1}{\mu} \frac{d\mu}{dT} \frac{E_F}{k_B}. \tag{3.1}$$

Here the values of $E_F$ was estimated for each $n$ by the formula $E_F = (\pi\hbar^2/m) n$.

The values of $A_{exp}$ are presented in Table 5. One can see that the theoretical and the experimental estimates of the effect (of the coefficients $A_{theor}$ and $A_{exp}$) are of the same order of magnitude but $A_{exp} > A_{theor}$ in (1.3 – 3.8) times for carriers of S - subbands and $A_{exp} < A_{theor}$ in (2.2 – 3.3) times for carriers of AS – subbands.

We empirically consider the renormalization of μ and/or $E_F$ in the (3.1) by introducing a correction factor $\alpha = A_{exp}/A_{theor}$, the values of which are shown in Table 5. For comparison, the results of Refs. 17,18 are also presented through the parameter $\alpha$ in Table 5. As can be seen the value of the correction factor $\alpha = A_{exp}/A_{theor}$ is closest to 1 (and to its value in Refs. 17, 18) for the sample DQW-3 with a maximum value of the parameter $k_F l \approx 30$.

Thus, the simple version of the theory [9] qualitatively describes our experimental data, grasping even such subtleties as the sign of the interaction effect in different subbands of size quantization.

Compared with [17,18], there are additional difficulties for both experimental and theoretical analysis of data in our systems. Uncertainties in the experimental results are related both to unavoidable measurement imprecisions and, especially, to the subtle procedure for separating the contributions of two types of carriers in double quantum wells.

On the other hand, we note the shortcomings of our theoretical estimates. First, accounting functions $f(k_B T\tau/\hbar)$ and $t(k_B T\tau/\hbar; \tilde{F}_0^\sigma)$ in (2.4) slightly reduces $A_{theor}$, thus increasing the discrepancy of the theory and experiment. Secondly, the expression (2.8) that connects $\tilde{F}_0^\sigma$ to $r_S$ is valid for $r_S < 1$ [9, 26, 27], while for our samples $r_S \cong (1 \div 3)$ (see Table 5), which introduces uncontrollable uncertainty into our estimates.

Thus, there is a qualitative agreement in the behavior of the contribution $\delta\sigma(T)$, observed in our experiments, and the EEI quantum correction to the conductivity of 2D system in the ballistic regime, calculated by Zala *et al.* [9]. The basic formulas of [9] unambiguously determine the sign of the derivative d[$\delta\sigma(T)$]/d$T$ in the wide temperature intervals from (10-20)K up to (45-70)K for investigated samples . We have found that a particular ("metallic" or "dielectric") type of the $\delta\sigma_{ee}^{ball}(T)$ behavior for a given substance depends only on the value of the carrier concentration.

## 5. Conclusions

We have measured the temperature and magnetic-field dependences of the longitudinal $\rho_{xx}(B,T)$ and Hall $\rho_{xy}(B,T)$ resistivities of GaAs/InGaAs/GaAs systems with a single or a double quantum well at $T= (1.8-100)$ K and $B$ up to 9T.

At $B$=0 a pronounced dielectric type of temperature dependence $\rho(T)$ ($d\rho/dT < 0$) occurs in a wide range of temperatures from $T \sim 10$K up to (45-75)K.

Analysis of our experimental data for the $\rho(T)$ at zero magnetic field allowed us both qualitatively and to a large extent quantitatively to explain the observed effects by the dominant contribution of the exchange EEI to the temperature dependence of the electron mobility in a ballistic regime.

A number of factors made it possible to observe clearly the "dielectric" behavior of the resistivity for the investigated series of samples over a temperature range corresponding to the quantum corrections from the EEI in the ballistic regime with the predominant role of the exchange term. They are:

the choice of the InGaAs material, which is suitable for microscopic parameters ($m$, $\kappa$), so that the concentration of the transition from the metallic behavior of $\delta\sigma^{ee}(T)$ to the dielectric one, $N_{cross} = 1.1 \cdot 10^{11}$ cm$^{-2}$, is in a range of values quite attainable experimentally; a successful set of electron concentrations in the systems under study: $n > N_{cross}$ for S subbands and $n < N_{cross}$ for AS subbands;

and, finally, (in contrast to GaAs), the predominantly short-range scattering potential, viz. the alloy scattering of electrons by In atoms as substitutional impurities.

## Acknowledgements

The research was carried out within the state assignment of FASO of Russia, theme "Electron" No. AAAA-A18-118020190098-5, supported in part by RFBR, projects Nos. 18-02-00172 (sample growth), 18-32-00382 (experiment), 18-02-00192 (theoretical support).